\documentclass[12pt]{spieman}
\usepackage{amsmath,amsfonts,amssymb}
\usepackage{graphicx,booktabs}
\usepackage{setspace,inputenc}
\usepackage[subfigure]{tocloft}
\usepackage{pgfplots,subfigure,rotating}
\usepackage{pdfpages,afterpage}
\usepackage{lineno}
\usepackage[bordercolor=white,backgroundcolor=gray!30,linecolor=black,colorinlistoftodos]{todonotes}

\title{Effects of Radiation Damage on the Optical Properties of Glass}

\author[a]{Federica Simonetto}
\author[b]{Matteo Marmonti}
\author[a,c]{Marco AC Potenza}
\affil[a]{University of Milan, Physics Department, via Celoria 16, Milan, Italy, 20133}
\affil[b]{Optec S.p.A., via Mantegna 34, Parabiago, Italy, 20015}
\affil[c]{CIMAINA, University of Milan, via Celoria, 16 I-20133 Milan, Italy}

\cftpagenumbersoff{figure}
\cftpagenumbersoff{table} 

\begin{document} 
\maketitle

\begin{abstract}
We study the optical properties of glass exposed to ionizing radiation, as it occurs in the space environment. 24 glass types have been considered, both space qualified and not space qualified. 72 samples (3 for each glass type) have been irradiated to simulate a total dose of 10krad and 30krad, imposed by a proton beam at KVI- Centre of Advanced Radiation Technology (Groeningen). Combining the information about stopping power and proton fluence, the time required to reproduce any given total dose in the real environment can be easily obtained. 
The optical properties, such as spectral transmission and light scattering have been measured before and after irradiation for each sample. 
Transmission has been characterized within the wavelength range 200 nm -- 1100 nm. Indications that systematical issues depend on the dopant or compiosition are found and described.  This work aims at extending the existing list of space--compliant glasses in terms of radiation damage. 
\end{abstract}


{\noindent \footnotesize\textbf{*}Federica Simonetto,  \linkable{federica.simonetto@unimi.it} }
{\noindent \footnotesize\textbf{*}Matteo Marmonti,  \linkable{mmarmonti@optec.eu} }
{\noindent \footnotesize\textbf{*}Marco Potenza,  \linkable{marco.potenza@unimi.it} }

\begin{spacing}{2}
\section{Introduction}
\label{sect:intro}
Space--born optical components require a well--characterized stability upon radiation damage in the space environment overall the expected lifetime of the instrument. 
Ionizing space radiation appreciably reduces optical transmission in many optical materials. The transmission loss can be severe depending on the material and the radiation dose \cite{fruit}. \\
Nowadays, a list of space--compliant glasses is available and dedicated measurements have been done to understand the expected degradation in space \cite{tammy}, \cite{manolis}. These materials are stabilized against transmittance losses by adding cerium (Ce) or other types of dopants \cite{kreidl}, \cite{stroud}. Although Cerium doping can affect the color of glass, it provides good stabilization against ionizing radiation damage. \\
Unfortunately, the stock availability of radiation hardened glasses is not always guaranteed. Moreover, their limited number is not enough to guarantee the optical designer to reach the high performances and tight tolerances required for new space optical systems. On the other hand, we also noticed that the orbit of many satellites (e.g. in Low Earth Observation orbits) can maintain the optical system within a limited radiation level where a conventional, non-radiation hard, glass could be safely adopted. However, Space Agencies are unwilling to use these materials due to missing experimental evidence.\\

In this work we study the influence of ionizing radiation on the optical properties of glass by reproducing the space conditions on ground, through a proper proton beam produced at the KVI- Centre of Advanced Radiation Technology (Groningen). Furthermore, numerical simulations about radiation absorption and the role of given chemical elements have been studied. The ultimate aim of this work is i) to expand the current list of usable materials and ii) to create a detailed test procedure as described below. 

We analyze  the changes in the spectral transmission and light scattering as a function of the proton fluency, Linear Energy Transfer (L.E.T.) and dose. Here we first-define and describe the approach to characterize the samples and we give a detailed description of the samples and physical quantities adopted. Then we describe the experiments performed and discuss the corresponding results.

\section{Methods}
\label{sect:meth}
\subsection{Transmittance}
Glass is usually transparent to visible light and some regions of the UV and IR ranges. Losses in transmittance occur due to internal absorption and reflection. The so-called internal transmittance $\tau_i$ can be modified by adding oxides of transition elements or small colloidal particles in the glass \cite{sanghera}, \cite{Potter:61}. The internal transmittance $\tau_i$ \cite{born2013principles} is a function of the sample thickness d: 
\begin{equation}
\tau_i = e^{-\alpha(\lambda) d}
\end{equation}
where $\alpha(\lambda)$ is the spectral absorption coefficient. The relation between spectral transmittance $\tau$ and spectral internal transmittance $\tau_i$ is:
\begin{equation}
\tau(\lambda) = \bigg(\dfrac{2n(\lambda)}{n^2 (\lambda) +1} \bigg) \tau_i (\lambda)
\end{equation}

Glass spectral transmission has been measured with DU800 Spectrophotometer \cite{manual}. Before and after irradiation, transmission have been measured six times for each sample, placed at six different positions in the spectrophotometer in order to estimate inhomogeneities. Results are averaged and, besides the spectral Transmission is given by integrating all over the spectrum.

\subsection{Scattering}
Light scattering spreads light from the incoming direction.  It can be caused by inhomogeneities in the propagation medium,  the presence of particles or defects within the medium as well as at the interface between two media \cite{jackson1999classical}, \cite{VdH}. In principle, light scattering from an ideal, defect--free crystalline bulk material is ultimately due to effects of anharmonicity within the lattice. Light wave transmission will be highly directional due to the typical optical anisotropy of crystalline substances, which depends on their symmetry group \cite{bohren2008absorption}. For example, the seven different crystalline forms of quartz silica (silicon dioxide, $SiO_2$) are clear, transparent. Hence the importance to qualify the effects induced by radiation within the glass in terms of light scattering in order to better inspect the structural changes. 

Light scattering has been measured by means of a Reflect 180S Goniophotometer \cite{refletwebsite}. 
More precisely, we measured the Bidirectional Scattering Distribution Function (BSDF). It is commonly adopted to provide the surface characterization of materials. Scattering is characterized in both reflection and transmission. The Bidirectional Reflectance Distribution Function (BRDF) is a function of four real variables that describes how much light power is reflected by an opaque surface. 
For a given incoming light direction, $\omega_i$, and outgoing direction, $\omega_r$, BRDF returns the ratio between the radiance reflected in direction $\omega_r$ and the irradiance incident on the surface in direction $\omega_i$. Units are, therefore, $1/sr$. BTDF (Bidirectional Transmission Distribution Function) is similarly defined on the basis of the transmission properties.

\subsection{Fluence, L.E.T. and dose}
In order to evaluate quantitatively the damage caused by radiation passing through a material, we introduce fluence, L.E.T. and absorbed dose. \\
Fluence is defined as the average number of particles impinging onto a unit surface (usually 1 $cm^2$). L.E.T. describes the effects of radiation as the energy transferred by the ionizing particles to the material per unit distance. The absorbed dose \cite{attix1986}, \cite{gre1981} represents the average energy released into matter per unit mass. Both L.E.T. and dose depend on the nature of the radiation and the material. In the International System of units, SI, dose is expressed in J/kg, or $gray$ (Gy). The common $rad$ corresponding to $10^{-2}$ Gy, is in CGS units. A closure exists for the relation between these quantities: if fluence, L.E.T. and exposure time are known, the total dose can be evaluated for different environments and conditions. We simulate the total dose using an open source tool called SPENVIS (\ref{appendix}). Moreover, we chose to consider only total dose according ESA specification\cite{ESA}.

\subsection{Samples and corresponding L.E.T. analysis}
The multitude of technical glasses can be roughly arranged in the following six groups, accordingly to their oxide composition (in weight fraction): Borosilicate (Non-Alkaline, Alkaline and High-Borate); Aluminosilicate (Alkaline and Alkali); Aluminoborosilicate;  Alkali-lead silicate; Alkali alkaline earth silicate (soda-lime glasses); LAS-glass-ceramics.
24 kinds of glass have been considered and studied with different setups that will be described later. For each material, three identical, cylindrical, flat slabs (20mm diameter and 5mm thickness) were used. 24 additional glasses have been used in one setup only. \\
Chemical composition is not known in details for all the samples. Weigh proportions about glasses known are reported in Tabs. \ref{tab:chemical_1}, \ref{tab:chemical_2}, \ref{tab:chemical_3} and \ref{tab:chemical_4}. 

\begin{table}[h]
\centering
\begin{tabular}{clcc}
\toprule
Glass Code	&	Chemical Name	&	Chemical Formula	&	Weight (\%)	\\
\midrule
802	&	Phosphorus pentoxide	& $	P_2O_5	$ & $	20-30	$ \\
	&	Titanium dioxide	& $	TiO_2	$ & $	10-20	$ \\
	&	Barium oxide	& $	BaO	$ & $	2-10	$ \\
	&	Silicon dioxide	& $	SiO_2	$ & $	0-2	$ \\
	&	Boron trioxide	& $	B_2O_3	$ & $	0-2	$ \\
	&	Antimony trioxide	& $	Sb_2O_3	$ & $	0-2	$ \\
\hline
804	&	Arsenic trioxide	& $	As_2O_3	$ & $	<1	$ \\
	&	Potassium oxide	& $	K_2O	$ & $	<1	$ \\
	&	Sodium oxide	& $	Na_2O	$ & $	<1	$ \\
	&	Lead oxide	& $	PbO	$ & $	70-80	$ \\
	&	Silicon dioxide	& $	SiO_2	$ & $	20-30	$ \\
\hline
805	&	Titanium dioxide	& $	TiO_2	$ & $	20-30	$ \\
	&	Silicon dioxide	& $	SiO_2	$ & $	20-30	$ \\
	&	Barium oxide	& $	BaO	$ & $	10-20	$ \\
	&	Zirconium oxide	& $	ZrO_2	$ & $	0-2	$ \\
	&	Antimony trioxide	& $	Sb_2O_3	$ & $	0-2	$ \\
\hline
806	&	Cerium oxide	& $	CeO_2	$ & $	<1	$ \\
	&	Potassium oxide	& $	K_2O	$ & $	1-10	$ \\
	&	Sodium oxide	& $	Na_2O	$ & $	<1	$ \\
	&	Lead oxide	& $	PbO	$ & $	70-80	$ \\
	&	Silicon dioxide	& $	SiO_2	$ & $	20-30	$ \\
\hline
807	&	Barium oxide	& $	BaO	$ & $	10-20	$ \\
	&	Calcium oxide	& $	CaO	$ & $	<1	$ \\
	&	Potassium oxide	& $	K_2O	$ & $	1-10	$ \\
	&	Sodium oxide	& $	Na_2O	$ & $	1-10	$ \\
	&	Niobium pentoxide	& $	Nb_2O_5	$ & $	10-20	$ \\
	&	Antimony trioxide	& $	Sb_2O_3	$ & $	<0.01	$ \\
	&	Silicon dioxide	& $	SiO_2	$ & $	20-30	$ \\
	&	Titanium dioxide	& $	TiO_2	$ & $	20-30	$ \\
\hline
808	&	Boron trioxide	& $	B_2O_3	$ & $	20-30	$ \\
	&	Barium oxide	& $	BaO	$ & $	<1	$ \\
	&	Lanthanum oxide	& $	La_2O_3	$ & $	40-50	$ \\
	&	Niobium pentoxide	& $	Nb_2O_5	$ & $	1-10	$ \\
	&	Antimony trioxide	& $	Sb_2O_3	$ & $	<0.01	$ \\
	&	Silicon dioxide	& $	SiO_2	$ & $	1-10	$ \\
	&	Yttrium oxide	& $	Y_2O_3	$ & $	1-10	$ \\
	&	Zinc oxide	& $	ZnO	$ & $	1-10	$ \\
	&	Zirconium oxide	& $	ZrO_2	$ & $	1-10	$ \\
\hline
810	&	Boron trioxide	& $	B_2O_3	$ & $	20-30	$ \\
	&	Zirconium oxide	& $	ZrO_2	$ & $	2-10	$ \\
	&	Silicon dioxide	& $	SiO_2	$ & $	2-10	$ \\
	&	Zinc oxide	& $	ZnO	$ & $	0-2	$ \\
	&	Antimony trioxide	& $	Sb_2O_3	$ & $	0-2	$ \\
\bottomrule
\end{tabular}
\caption{Chemical weight in percentage about glasses from 802 to 810.}
\label{tab:chemical_1}
\end{table}

\begin{table}[h]
\centering
\begin{tabular}{clcc}
\toprule
Glass Code	&	Chemical Name	&	Chemical Formula	&	Weight (\%)	\\
\midrule
812	&	Silicon dioxide	& $	SiO_2	$ & $	30-40	$ \\
	&	Boron trioxide	& $	B_2O_3	$ & $	10-20	$ \\
	&	Zirconium oxide	& $	ZrO_2	$ & $	2-10	$ \\
	&	Zinc oxide	& $	ZnO	$ & $	2-10	$ \\
	&	Calcium oxide	& $	CaO	$ & $	2-10	$ \\
	&	Barium oxide	& $	BaO	$ & $	0-2	$ \\
	&	Antimony trioxide	& $	Sb_2O_3	$ & $	0-2	$ \\
\hline
813	&	Boron trioxide	& $	B_2O_3	$ & $	<1	$ \\
	&	Calcium oxide	& $	CaO	$ & $	1-10	$ \\
	&	Potassium oxide	& $	K_2O	$ & $	1-10	$ \\
	&	Sodium oxide	& $	Na_2O	$ & $	1-10	$ \\
	&	Antimony trioxide	& $	Sb_2O_3	$ & $	<1	$ \\
	&	Silicon dioxide	& $	SiO_2	$ & $	50-60	$ \\
	&	Strontium oxide	& $	SrO	$ & $	1-10	$ \\
	&	Titanium dioxide	& $	TiO_2	$ & $	20-30	$ \\
\hline
814	&	Boron trioxide	& $	B_2O_3	$ & $	1-10	$ \\
	&	Calcium oxide	& $	CaO	$ & $	<1	$ \\
	&	Potassium oxide	& $	K_2O	$ & $	1-10	$ \\
	&	Sodium oxide	& $	Na_2O	$ & $	1-10	$ \\
	&	Antimony trioxide	& $	Sb_2O_3	$ & $	<1	$ \\
	&	Silicon dioxide	& $	SiO_2	$ & $	50-60	$ \\
	&	Titanium dioxide	& $	TiO_2	$ & $	10-20	$ \\
\hline
815	&	Cerium oxide	& $	CeO_2	$ & $	1-10	$ \\
	&	Potassium oxide	& $	K_2O	$ & $	1-10	$ \\
	&	Sodium oxide	& $	Na_2O	$ & $	1-10	$ \\
	&	Lead oxide	& $	PbO	$ & $	40-50	$ \\
	&	Silicon dioxide	& $	SiO_2	$ & $	40-50	$ \\
\hline
816	&	Silicon dioxide	& $	SiO_2	$ & $	30-40	$ \\
	&	Boron trioxide	& $	B_2O_3	$ & $	10-20	$ \\
	&	Zirconium oxide	& $	ZrO_2	$ & $	2-10	$ \\
	&	Barium oxide	& $	BaO	$ & $	2-10	$ \\
	&	Tantalum oxide	& $	Ta_2O_5	$ & $	2-10	$ \\
	&	Zinc oxide	& $	ZnO	$ & $	2-10	$ \\
	&	Calcium oxide	& $	CaO	$ & $	0-2	$ \\
	&	Antimony trioxide	& $	Sb_2O_3	$ & $	0-2	$ \\
\bottomrule
\end{tabular}
\caption{Chemical weight in percentage about glasses from 812 to 816.}
\label{tab:chemical_2}
\end{table}

\begin{table}[h]
\centering
\begin{tabular}{clcc}
\toprule
Glass Code	&	Chemical Name	&	Chemical Formula	&	Weight (\%)	\\
\midrule
817	&	Boron trioxide	& $	B_2O_3	$ & $	30-40	$ \\
	&	Barium oxide	& $	BaO	$ & $	1-10	$ \\
	&	Calcium oxide	& $	CaO	$ & $	10-20	$ \\
	&	Cerium oxide	& $	CeO_2	$ & $	1-10	$ \\
	&	Lanthanum oxide	& $	La_2O_3	$ & $	20-30	$ \\
	&	Magnesium oxide	& $	MgO	$ & $	1-10	$ \\
	&	Silicon dioxide	& $	SiO_2	$ & $	1-10	$ \\
	&	Zinc oxide	& $	ZnO	$ & $	1-10	$ \\
	&	Zirconium oxide	& $	ZrO_2	$ & $	1-10	$ \\
\hline
818	&	Barium oxide	& $	BaO	$ & $	40-50	$ \\
	&	Boron trioxide	& $	B_2O_3	$ & $	20-30	$ \\
	&	Silicon dioxide	& $	SiO_2	$ & $	10-20	$ \\
	&	Zirconium oxide	& $	ZrO_2	$ & $	2-10	$ \\
	&	Aluminium oxide	& $	Al_2O_3	$ & $	0-2	$ \\
	&	Antimony trioxide	& $	Sb_2O_3	$ & $	0-2	$ \\
\hline
819	&	Barium fluoride	& $	BaF_2	$ & $	30-40	$ \\
	&	Phosphorus pentoxide	& $	P_2O_5	$ & $	20-30	$ \\
	&	Barium oxide	& $	BaO	$ & $	20-30	$ \\
	&	Aluminium oxide	& $	Al_2O_3	$ & $	2-10	$ \\
	&	Aluminum fluoride	& $	AlF_3	$ & $	2-10	$ \\
	&	Calcium fluoride	& $	CaF_2	$ & $	0-2	$ \\
\hline
820	&	Phosphorus pentoxide	& $	P_2O_5	$ & $	40-50	$ \\
	&	Barium oxide	& $	BaO	$ & $	30-40	$ \\
	&	Calcium oxide	& $	CaO	$ & $	2-10	$ \\
	&	Boron trioxide	& $	B_2O_3	$ & $	2-10	$ \\
	&	Zinc oxide	& $	ZnO	$ & $	0-2	$ \\
	&	Tungsten oxide	& $	WO_3	$ & $	0-2	$ \\
	&	Antimony trioxide	& $	Sb_2O_3	$ & $	0-2	$ \\
\hline
821	&	Boron trioxide	& $	B_2O_3	$ & $	1-10	$ \\
	&	Calcium oxide	& $	CaO	$ & $	1-10	$ \\
	&	Cerium oxide	& $	CeO_2	$ & $	1-10	$ \\
	&	Potassium oxide	& $	K_2O	$ & $	10-20	$ \\
	&	Sodium oxide	& $	Na_2O	$ & $	1-10	$ \\
	&	Lead oxide	& $	PbO	$ & $	1-10	$ \\
	&	Silicon dioxide	& $	SiO_2	$ & $	60-70	$ \\
	&	Zinc oxide	& $	ZnO	$ & $	1-10	$ \\
\bottomrule
\end{tabular}
\caption{Chemical weight in percentage about glasses from 817 to 821.}
\label{tab:chemical_3}
\end{table}

\begin{table}[h]
\centering
\begin{tabular}{clcc}
\toprule
Glass Code	&	Chemical Name	&	Chemical Formula	&	Weight (\%)	\\
\midrule
822	&	Barium oxide	& $	BaO	$ & $	40-50	$ \\
	&	Silicon dioxide	& $	SiO_2	$ & $	30-40	$ \\
	&	Boron trioxide	& $	B_2O_3	$ & $	10-20	$ \\
	&	Aluminium oxide	& $	Al_2O_3	$ & $	2-10	$ \\
	&	Antimony trioxide	& $	Sb_2O_3	$ & $	0-2	$ \\
	&	Titanium dioxide	& $	TiO_2	$ & $	0-2	$ \\
\hline
823	&	Aluminium oxide	& $	Al_2O_3	$ & $	1-10	$ \\
	&	Boron trioxide	& $	B_2O_3	$ & $	40-50	$ \\
	&	Calcium oxide	& $	CaO	$ & $	1-10	$ \\
	&	Potassium oxide	& $	K_2O	$ & $	<1	$ \\
	&	Lithium oxide	& $	Li_2O	$ & $	1-10	$ \\
	&	Sodium oxide	& $	Na_2O	$ & $	1-10	$ \\
	&	Niobium pentoxide	& $	Nb_2O_5	$ & $	<1	$ \\
	&	Antimony trioxide	& $	Sb_2O_3	$ & $	<1	$ \\
	&	Silicon dioxide	& $	SiO_2	$ & $	10-20	$ \\
	&	Tantalum oxide	& $	Ta_2O_5	$ & $	10-20	$ \\
	&	Zinc oxide	& $	ZnO	$ & $	1-10	$ \\
	&	Zirconium oxide	& $	ZrO_2	$ & $	1-10	$ \\
\hline
826	&	Boron trioxide	& $	B_2O_3	$ & $	10-20	$ \\
	&	Barium oxide	& $	BaO	$ & $	1-10	$ \\
	&	Cerium oxide	& $	CeO_2	$ & $	1-10	$ \\
	&	Potassium oxide	& $	K_2O	$ & $	1-10	$ \\
	&	Sodium oxide	& $	Na_2O	$ & $	1-10	$ \\
	&	Silicon dioxide	& $	SiO_2	$ & $	60-70	$ \\
	&	Zinc oxide	& $	ZnO	$ & $	<1	$ \\
\hline
827	&	Boron trioxide	& $	B_2O_3	$ & $	10-20	$ \\
	&	Barium oxide	& $	BaO	$ & $	1-10	$ \\
	&	Calcium oxide	& $	CaO	$ & $	<1	$ \\
	&	Chlorine	& $	Cl	$ & $	<1	$ \\
	&	Potassium oxide	& $	K_2O	$ & $	1-10	$ \\
	&	Sodium oxide	& $	Na_2O	$ & $	10-20	$ \\
	&	Antimony trioxide	& $	Sb_2O_3	$ & $	<1	$ \\
	&	Silicon dioxide	& $	SiO_2	$ & $	60-70	$ \\
	&	Titanium dioxide	& $	TiO_2	$ & $	<1	$ \\
\hline
828	&	Strontium fluoride	& $	SrF_2	$ & $	20-30	$ \\
	&	Phosphorus pentoxide	& $	P_2O_5	$ & $	20-30	$ \\
	&	Barium fluoride	& $	BaF_2	$ & $	10-20	$ \\
	&	Calcium fluoride	& $	CaF_2	$ & $	10-20	$ \\
	&	Aluminum fluoride	& $	AlF_3	$ & $	10-20	$ \\
	&	Magnesium fluoride	& $	MgF_2	$ & $	2-10	$ \\
	&	Aluminium oxide	& $	Al_2O_3	$ & $	2-10	$ \\
	&	Barium oxide	& $	BaO	$ & $	0-2	$ \\
\bottomrule
\end{tabular}
\caption{Chemical weight in percentage about glasses from 822 to 828.}
\label{tab:chemical_4}
\end{table}

For our purpose, we use the chemical composition of 11 types of glasses from literature \cite{bass}. L.E.T. has been evaluated through the open--source software called "Stopping and Range of Ions in Matter" (SRIM). It allows to estimate the L.E.T. for different proton energies by introducing the composition of the material \cite{edw75}. 
In this way we can analyze the dependence between L.E.T., density and proton energy and to attempt to understand the effect of a specific chemical element. 

We report the results of simulation for particular doses, such those typically encountered in space missions. L.E.T. values and times have been combined according to:
\begin{equation}
\label{eq:time}
time = \dfrac{Fluence}{Flux} = \dfrac{Dose}{L.E.T.} \cdot \dfrac{1}{Flux}
\end{equation}

\begin{figure}[!h]
\centering
\includegraphics[scale=0.5]{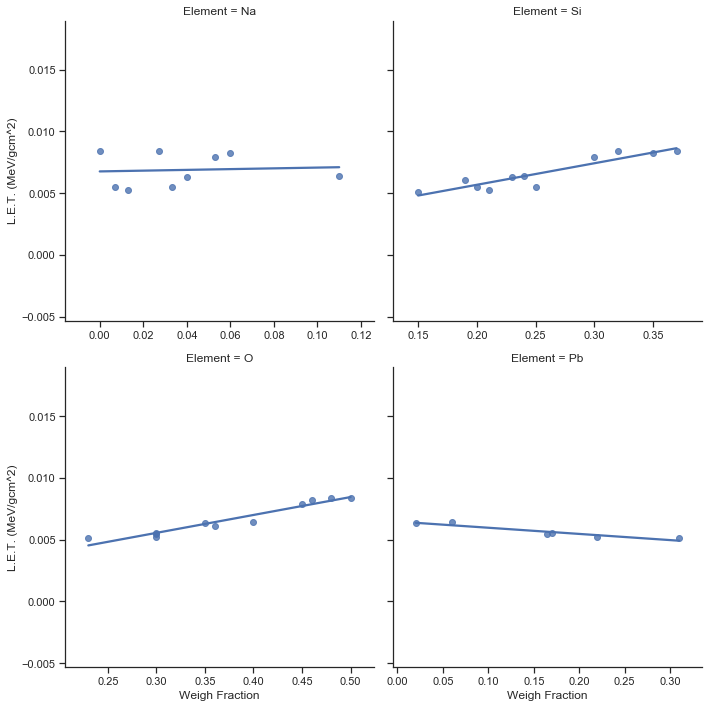}
\caption{L.E.T. plotted against the weight fraction of four elements: Sodium, Silicon, Oxigen, Lead. Each glass is represented by a circle in this plot, the line just shows the best linear fit as a guide to the eye.}
\label{fig:correlation}
\end{figure}

We performed a preliminary analysis to extract the influence on the L.E.T. of given chemical elements composing a glass. 
The mass fraction of a given element, the L.E.T. value and the glass average density have been considered. In Fig.\ref{fig:correlation} L.E.T. is plotted against the mass fraction of the most common elements. Glasses with high fraction of Lead mainly correspond to low L.E.T. values, therefore small radiation damage. On the other hand, glasses containing Silicon and Oxygen show higher L.E.T. values, that implies higher damage levels. thus, L.E.T. appears to be correlated to the presence of Silicon and Oxygen, anti-correlated to Lead.

\begin{figure}[!h]
\centering
\includegraphics[scale=0.7]{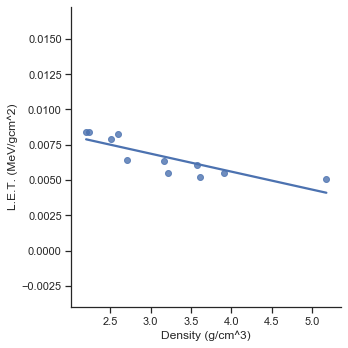}
\caption{L.E.T. is expressed in $MeV/gcm^2$, density in $g/cm^3$. Each glass is represented by a circle in this plot, the line just shows the best linear fit as a guide to the eye.}
\label{fig:density}
\end{figure}

In Fig \ref{fig:density} L.E.T./density trend is reported. The higher the density, the lower the L.E.T. This leads to the conclusion that lower density glasses will be damage more by ionizing energy.

Finally, when fluence is known it is possible to evaluate irradiation time to impose a given dose. Assuming a fluence of $10^8$ particles/s $cm^2$, to give a dose of 10krad and 30krad we need irradiation times given by equation \ref{eq:time}.

The corresponding results are reported in Tab \ref{tab:time10} for 10krad, in Tab \ref{tab:time30} for 30krad. \\

\begin{table}[h]
\centering
\begin{tabular}{lcccccc}
\toprule
\textbf{Glass} &  Fluence & Fluence & Fluence & Time & Time & Time \\
 & at 65MeV & at 100MeV & at 150MeV & for 65MeV & for 100MeV & for 150MeV \\
 & p/$cm^2$ & p/$cm^2$ & p/$cm^2$& s & s & s \\
\midrule
Fused Silica & 7,45E+10 & 1,02E+11 & 1,36E+11 & 744,71 & 1018,14 & 1359,36 \\
Borosilicate & 7,89E+10 & 1,09E+11 & 1,45E+11 & 789,35 & 1089,96 & 1452,35 \\
Crown & 7,57E+10 & 1,05E+11 & 1,40E+11 & 757,27 & 1046,82 & 1395,84 \\
Crown Flint & 9,74E+10 & 1,54E+11 & 2,03E+11 & 973,67 & 1540,90 & 2028,00 \\
Dense Ba Crown & 1,03E+11 & 1,61E+11 & 2,12E+11 & 1027,52 & 1614,64 & 2121,03 \\
Dense Ba Flint & 1,14E+11 & 1,55E+11 & 2,04E+11 & 1139,71 & 1549,32 & 2039,27 \\
Dense Flint & 1,22E+11 & 1,66E+11 & 2,17E+11 & 1223,48 & 1656,63 & 2174,22 \\
Flint & 1,19E+11 & 1,61E+11 & 2,12E+11 & 1190,80 & 1614,64 & 2121,03 \\
Light Flint & 1,13E+11 & 1,54E+11 & 2,03E+11 & 1133,30 & 1540,90 & 2028,00 \\
Light Flint Ba & 9,87E+10 & 1,35E+11 & 1,79E+11 & 987,22 & 1351,12 & 1788,08 \\
Pyrex & 7,44E+10 & 1,03E+11 & 1,37E+11 & 744,18 & 1029,73 & 1374,33 \\
\bottomrule
\end{tabular}
\caption{Irradiation time needed to impose a dose of 10 krad, supposing a flux of $10^8$ particles/s $cm^2$ and L.E.T.  estimated for three energies.}
\label{tab:time10}
\end{table}

\begin{table}[h]
\centering
\begin{tabular}{lcccccc}
\toprule
\textbf{Glass} &  Fluence & Fluence & Fluence & Time & Time & Time \\
 & at 65MeV & at 100MeV & at 150MeV & for 65MeV & for 100MeV & for 150MeV \\
 & p/$cm^2$ & p/$cm^2$ & p/$cm^2$& s & s & s \\
\midrule
Fused Silica	&	2,23E+11	&	3,05E+11	&	4,08E+11	&	2234,14	&	3054,41	&	4078,09	\\
Borosilicate	&	2,37E+11	&	3,27E+11	&	4,36E+11	&	2368,06	&	3269,88	&	4357,05	\\
Crown	&	2,27E+11	&	3,14E+11	&	4,19E+11	&	2271,82	&	3140,46	&	4187,52	\\
Crown Flint	&	2,92E+11	&	4,62E+11	&	6,08E+11	&	2921,01	&	4622,71	&	6084,01	\\
Dense Ba Crown	&	3,08E+11	&	4,84E+11	&	6,36E+11	&	3082,57	&	4843,92	&	6363,10	\\
Dense Ba Flint	&	3,42E+11	&	4,65E+11	&	6,12E+11	&	3419,13	&	4647,95	&	6117,80	\\
Dense Flint	&	3,67E+11	&	4,97E+11	&	6,52E+11	&	3670,44	&	4969,90	&	6522,67	\\
Flint	&	3,57E+11	&	4,84E+11	&	6,36E+11	&	3572,41	&	4843,92	&	6363,10	\\
Light Flint	&	3,40E+11	&	4,62E+11	&	6,08E+11	&	3399,89	&	4622,71	&	6084,01	\\
Light Flint Ba	&	2,96E+11	&	4,05E+11	&	5,36E+11	&	2961,66	&	4053,37	&	5364,25	\\
Pyrex	&	2,23E+11	&	3,09E+11	&	4,12E+11	&	2232,55	&	3089,18	&	4122,98	\\

\bottomrule
\end{tabular}
\caption{Irradiation time needed to impose a dose of 30 krad with a flux of $10^8$ particles/s $cm^2$ and L.E.T.  estimated for three energies.}
\label{tab:time30}
\end{table}

Irradiation took place at KVI-CART in Groningen \cite{kviwebsite}. The standard irradiation section has a diameter of $70$ mm and homogeneity of better than $\pm 3\%$. Larger fields (up to 110 to 140mm in diameter) can be realized with homogeneities better than $\pm 10\%$ and $\pm 25\%$, respectively. The samples have been fixed on aluminum holders with adhesive tape. Three different setups have been arranged for the tests: 
\begin{itemize}
\item Setup a: one sample for each type of glass, irradiated with a dose of 30krad; 
\item Setup b: two samples, one for each type of glasses and an additional one (825a), irradiated with a dose of 30krad;
\item Setup c: one sample for each type of glass, irradiated with a dose of 10krad. . 
\end{itemize}

To impose the same dose at different glasses, samples have been divided into groups, with proper given irradiation times. Glasses have also been divided in two categories: simil-Flint and simil-Crown. In this way, different times of irradiation have been evaluated. Results are reported in Tab. \ref{tab:glasstable}.

\begin{table}[!h]
\centering
\begin{tabular}{cc|cc|cc}
\toprule
& Setup a &  & Setup b & & Setup c\\
3400 s & 2300 s & 3400 s & 2300 s & 1100 s & 800 s \\
\midrule
802a & 817a & 802b & 817b & 802c & 817c \\  
804a & 818a & 804b & 818b & 804c & 818c \\ 
805a & 819a & 805b & 819b & 805c & 819c \\ 
806a & 820a & 806b & 820b & 806c & 820c \\  
807a & 821a & 807b & 821b & 807c & 821c \\  
808a & 822a & 808b & 822b & 808c & 822c \\ 
810a & 823a & 810b & 823b & 810c & 823c \\ 
811a & 826a & 811b & 826b & 811c & 826c \\
812a & 827a & 812b & 827b & 812c & 827c \\  
813a & 828a & 813b & 828b & 813c & 828c \\ 
814a &  & 814b &  & 814c &  \\ 
815a &  & 815b &  & 815c &  \\
816a &  & 816b &  & 816c &  \\ 
825a &  & 825b &  & 825c &  \\
\bottomrule
\end{tabular}
\caption{Irradiation times for different types of glass. For each setup, the right column contanins the flint glasses and the left column contains crown glasses. Notice that the letter indicates the adopted setup. }
\label{tab:glasstable}
\end{table}

\section{Experimental data and analysis}
\label{sect:analysis}
\subsection{Transmission analysis}
\begin{figure}[h!]
\centering
\caption{\textbf{Glass 806}: transmission before and after irradiation. (a): Glass type \emph{a}: irradiation setup for 30krad. (b): Glass type \emph{b}: irradiation setup for 30krad, with glass 825a. (c): Glass type \emph{c}: irradiation setup for 10krad. (d): Ratio between transmission after irradiation and before irradiation.}
\includegraphics[scale=0.55]{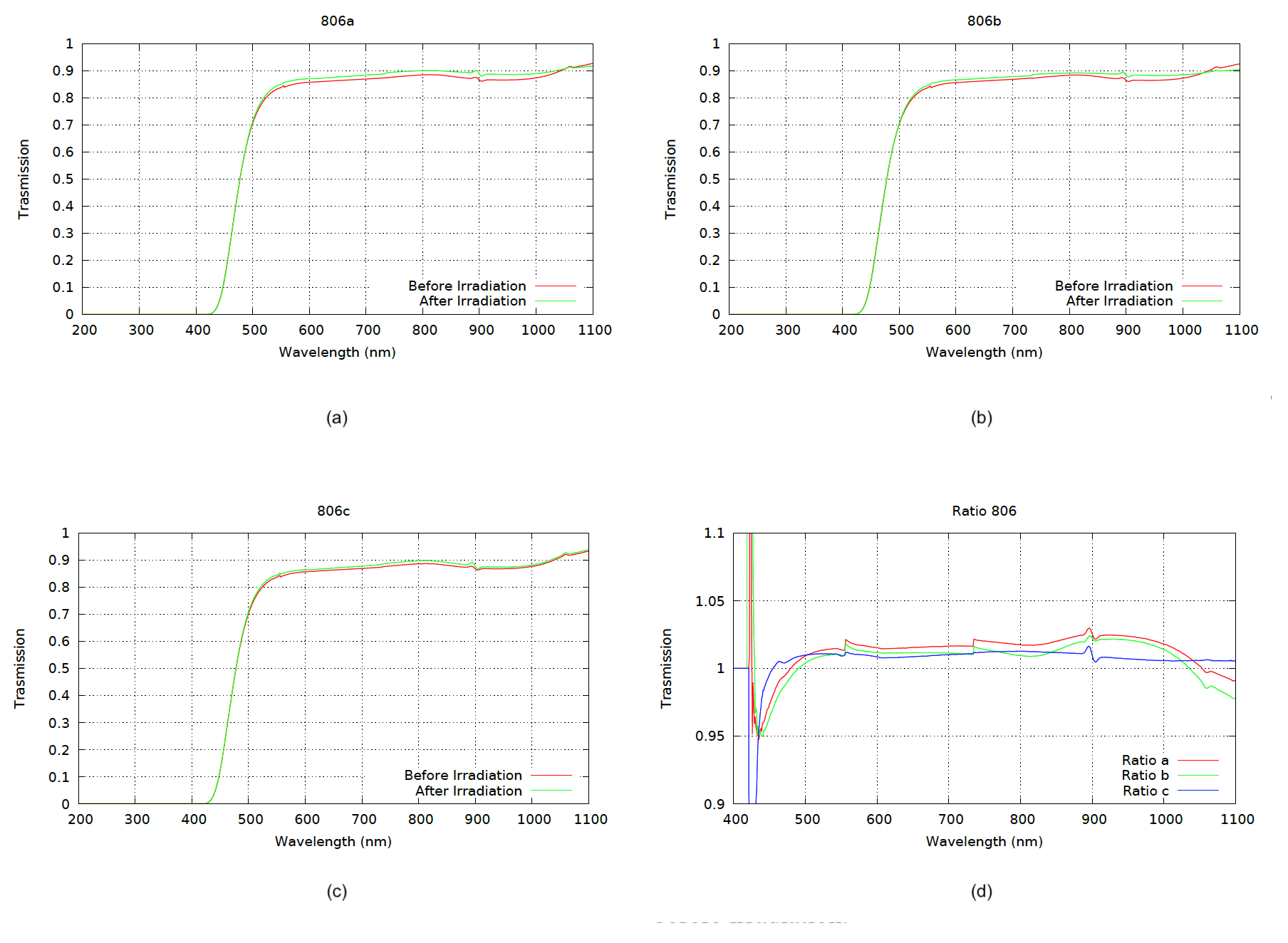}
\label{fig:t806}
\end{figure}

\begin{figure}[h!]
\centering
\caption{\textbf{Glass 802}: transmission before and after irradiation. (a): Glass type \emph{a}: irradiation setup for 30krad. (b): Glass type \emph{b}: irradiation setup for 30krad, with glass 825a. (c): Glass type \emph{c}: irradiation setup for 10krad. (d): Ratio between transmission after irradiation and before irradiation.}
\includegraphics[scale=0.55]{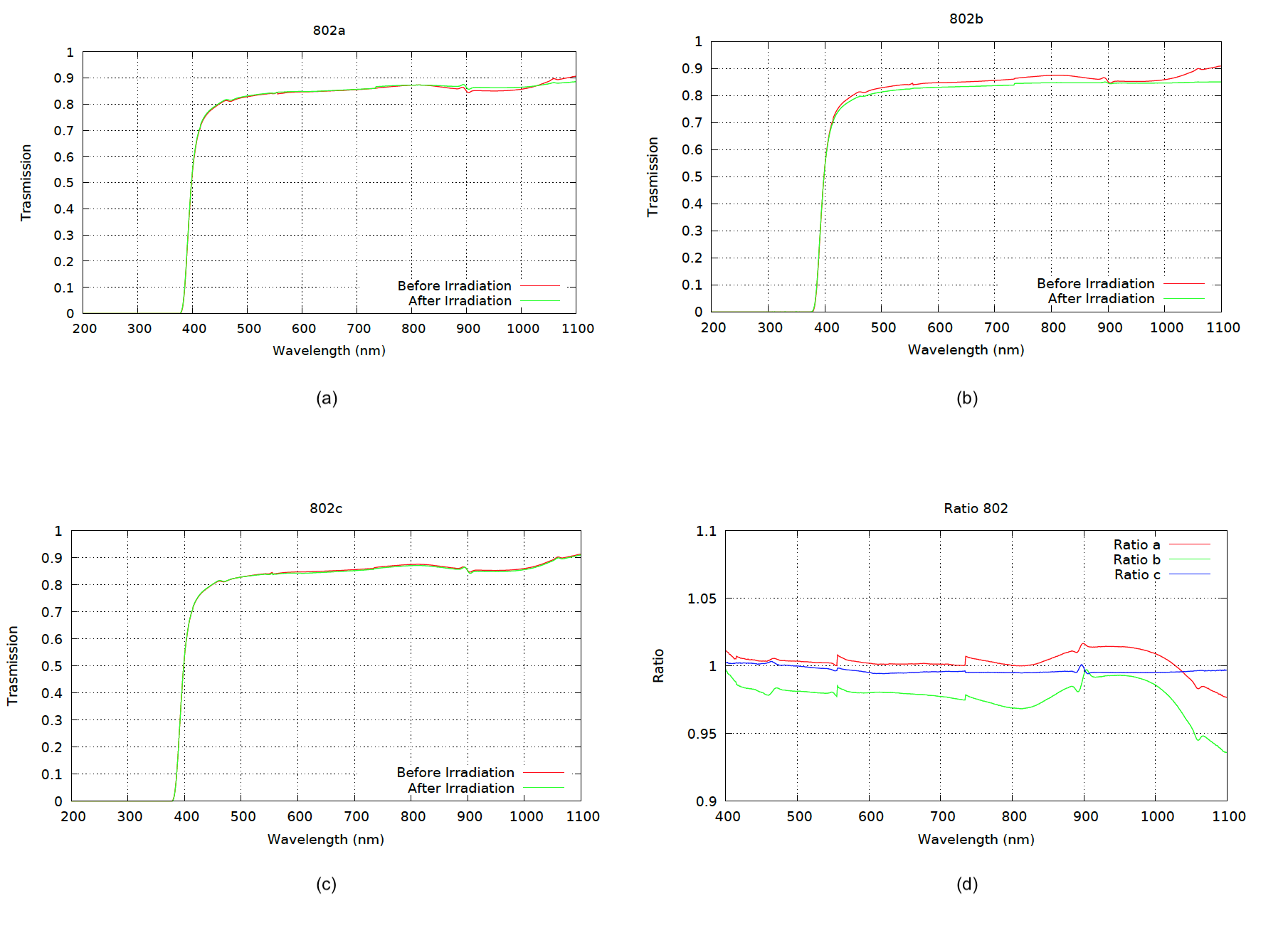}
\label{fig:t802}
\end{figure}

\begin{figure}[h!]
\centering
\caption{\textbf{Glass 810}: transmission before and after irradiation. (a): Glass type \emph{a}: irradiation setup for 30krad. (b): Glass type \emph{b}: irradiation setup for 30krad, with glass 825a. (c): Glass type \emph{c}: irradiation setup for 10krad. (d): Ratio between transmission after irradiation and before irradiation.}
\includegraphics[scale=0.55]{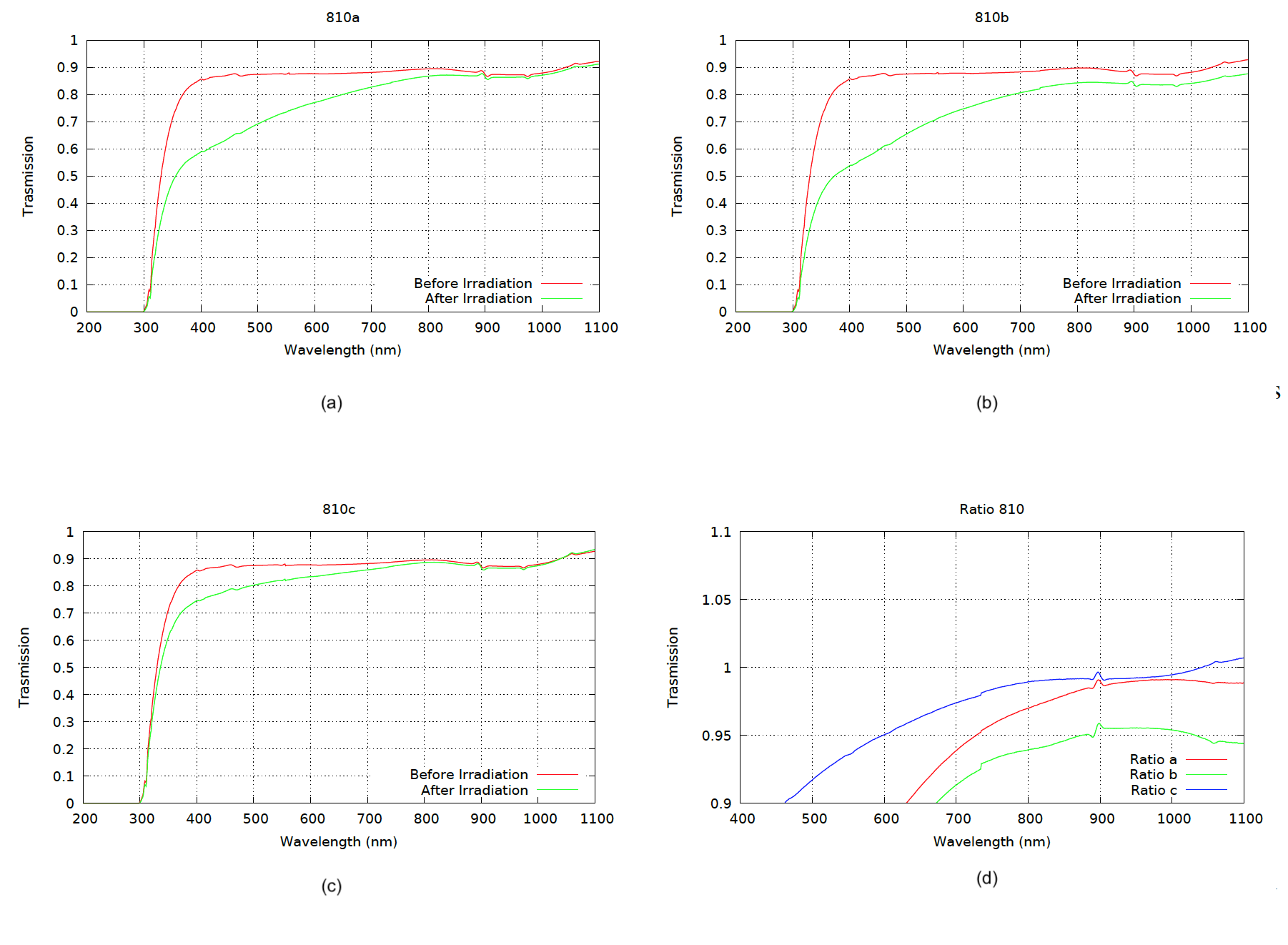}
\label{fig:t810}
\end{figure}

\begin{figure}[h!]
\centering
\caption{\textbf{Glass 820}: transmission before and after irradiation. (a): Glass type \emph{a}: irradiation setup for 30krad. (b): Glass type \emph{b}: irradiation setup for 30krad, with glass 825a. (c): Glass type \emph{c}: irradiation setup for 10krad. (d): Ratio between transmission after irradiation and before irradiation.}
\includegraphics[scale=0.55]{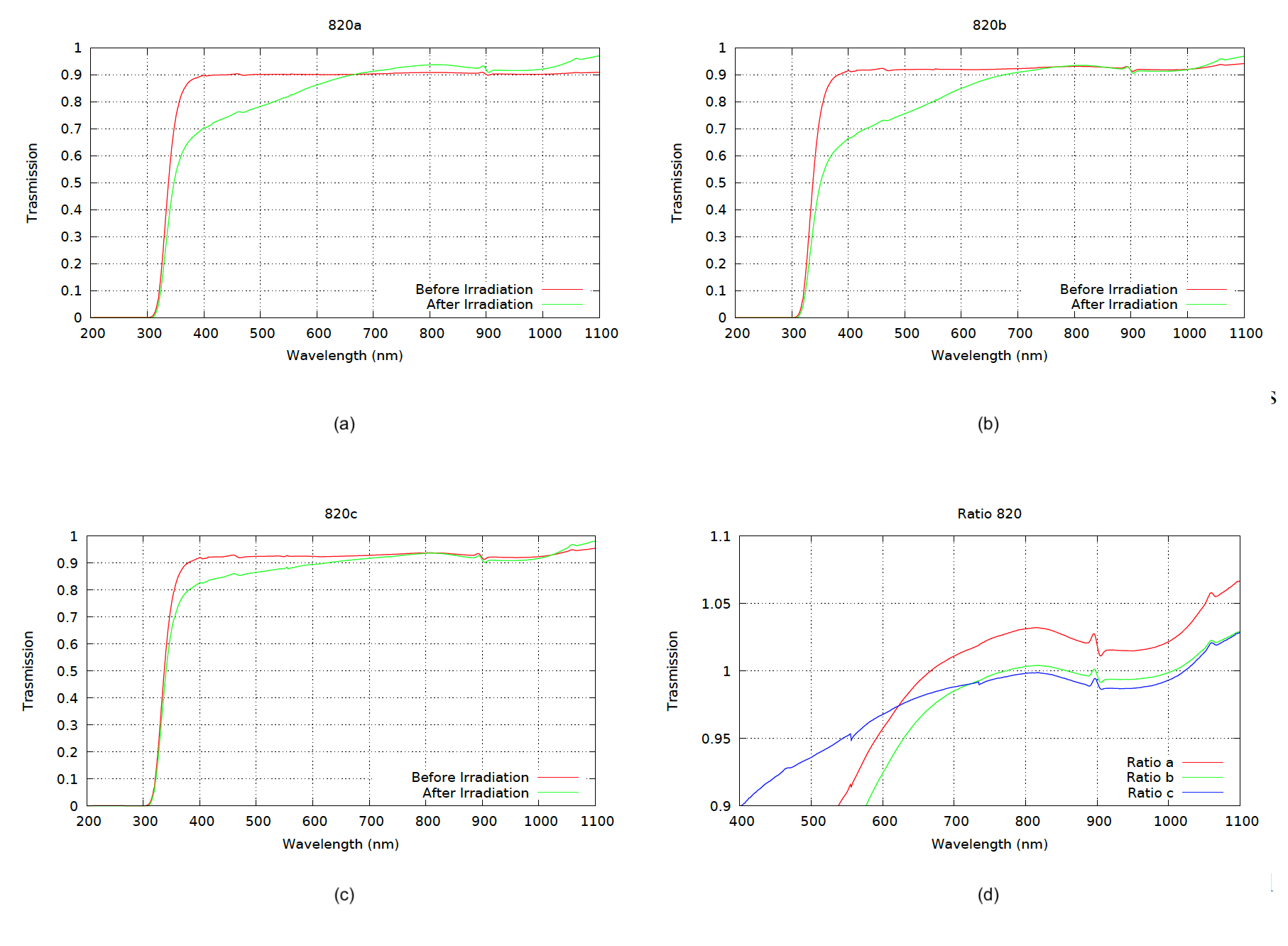}
\label{fig:t820}
\end{figure}

Hereinafter some plots are reported as examples of the experimental results obtained with our transmission analysis. Four plots are shown for each sample: three of them show transmission before and after irradiation for cases a, b and c; the last shows the ratio between two transmission values for the same wavelength (transmission after irradiation /transmission before irradiation). Sample 806 (Fig.\ref{fig:t806}) is an example of results obtained with a space compliant glass; 802 (Fig.\ref{fig:t802}) is an example of no--space compliant glass with transmission changing less than 5\%; 820 (Fig.\ref{fig:t820}) shows a transmission change larger than 5\%; 810 (Fig.\ref{fig:t810}) is an example of Lantanium glass. 

As a general conclusion of our analysis, we preliminarily catch the reader's attention to the following results: 

\begin{itemize}
\item even before irradiation, transmission is not 100\% because our glasses are not coated;
\item if the transmission changes are less than 5\%, the glass is considered as acceptable.For the present study a tolerance of 5\% has been considered acceptable since the optical telescopes this study has been conducted for had the following requirements:
\begin{itemize}
\item Transmission BOL (Begin of Life) is greater than 90\%
\item Transmission EOL (End of Life) is greater than 40\%
\end{itemize}
Assuming the optical system is composed by 10 lenses, a decrease of 5\% for each lens, means that the overall transmission EOL (i.e. once the total radiation dose has been applied) is still on spec (over 55\%). Space projects require a minimum EOL transmission due to glass coating and optical surface.  For space missions taken into account in this work, the transmission change accepted is 8\%, so we decided to decrease that percentage to 5 in order to expand the analysis to a wider projects range.

\item The samples that show a transmission change smaller than 5\% are: 802, 805 (limited to wavelengths larger that $\lambda=500 nm$) and 807. 

\item For glass 806, a slight increase seems to be there, as we have verified by repeating the measurements many times: transmission measurements always show an increase after irradiation. We can reasonably exclude experimental errors.
 
\item Other glasses show a transmission change smaller than 5\% only for one setup, for example 812, 813, 814, 823 (only above $\lambda= 600 nm$) and 827 (only above $\lambda= 500 nm)$ are acceptable only for setup c. 804 is acceptable for setup c and is acceptable only between $\lambda=500 nm$ and $\lambda=1100 nm$ for setup a and b;

\item Glasses with transmission changes larger than 5\% are:  816, 818, 820 and 822. These glasses are considered not suitable for space applications; 

\item Two glasses present non--monotonic behavior: 819 and 828. Both contain Fluorine, presenting low refractive index and high Abbe number;

\item 808, 810 and 811 present the same trend for transmission data before and after irradiation. They contain Lantanium.
\end{itemize}

Finally, three tables are presented, one for each experimental setup $a$ (Tables \ref{tab:Ttable_a1} and \ref{tab:Ttable_a2} ), $b$ (Tables \ref{tab:Ttable_b1} and \ref{tab:Ttable_b2}) and $c$ (Tables \ref{tab:Ttable_c1} and \ref{tab:Ttable_c2}). Tables show transmission variation at certain wavelength, chosen among glasses catalog standard wavelengths. For each value, an error of 0,00001 is considered, due to instrumental limitations.\\

\subsection{Scattering analysis}
Scattering measurements have been performed before and after irradiation at LightTec, France, for the glasses irradiated in the setup for 30krad. Measurements have been taken by illuminating the samples with collimated white light at an incidence angle of $10^{\circ}$ and $30^{\circ}$. A step of 0.1 degrees has been used and data have been processed to extrapolate the scattering function in reflection (BRDF) and in transmission (BTDF). \\
Data have been normalized at the intensity of the incoming light beam. The peaks of both the scattering functions before and after irradiation have been overlapped to compare the widths. Finally, the ratio between the two functions has been calculated to evidence the changes.\\
The small shift of the curves along the horizontal axis (angle) is due to an instrumental error. Results are presented for each glass: plots show BRDF, BTDF and FWHM, at $30^{\circ}$, before and after irradiation (Figures at page \pageref{fig:806} and thereafter).

Comparing the information obtained from the scattering analysis, the following conclusions can be drawn:
\begin{itemize}
\item generally speaking, the scattering of both transmitted and reflected light increases after irradiation. 
This could be due to the increase of either surface roughness or internal inhomogeneities, or both. 
This brings to an increase of the angular spread of the scattered light. 
\item Lead glass presents moderate changes in terms of light scattering.
\item glass containing Zinc or Boron exhibits remarkable increase of light scattering.
\end{itemize}
Finally, the following table summarizes all the results obtained for the light scattering analysis. The percentage variation between FWHM evaluated before and after irradiation is reported for BRDF and BTDF. Errors are evaluated in terms of the standard deviation.\\

\begin{table}[!h]
\centering
\begin{tabular}{c|cc|cc}
\toprule
Glass Code & BRDF &Dev & BTDF & Dev\\
& $\Delta$ FWHM & STD & $\Delta$ FWHM & STD \\
\midrule
802	&	0,26	&	0,03	&	0,25	&	0,03	\\
804	&	0,24	&	0,02	&	0,23	&	0,02	\\
805	&	0,22	&	0,02	&	0,23	&	0,02	\\
806	&	0,23	&	0,02	&	0,23	&	0,02	\\
807	&	0,27	&	0,03	&	0,23	&	0,02	\\
808	&	0,26	&	0,03	&	0,26	&	0,03	\\
810	&	0,23	&	0,02	&	0,23	&	0,02	\\
811	&	0,23	&	0,02	&	0,22	&	0,02	\\
812	&	0,22	&	0,02	&	0,22	&	0,02	\\
813	&	0,22	&	0,02	&	0,22	&	0,02	\\
814	&	0,24	&	0,02	&	0,22	&	0,02	\\
815	&	0,23	&	0,02	&	0,23	&	0,02	\\
816	&	0,29	&	0,03	&	0,28	&	0,03	\\
817	&	0,26	&	0,03	&	0,23	&	0,02	\\
818	&	0,24	&	0,02	&	0,23	&	0,02	\\
819	&	0,25	&	0,03	&	0,23	&	0,02	\\
820	&	0,23	&	0,02	&	0,22	&	0,02	\\
821	&	0,22	&	0,02	&	0,22	&	0,02	\\
822	&	0,24	&	0,02	&	0,23	&	0,02	\\
823	&	0,28	&	0,03	&	0,24	&	0,02	\\
825	&	0,20	&	0,02	&	0,23	&	0,02	\\
826	&	0,22	&	0,02	&	0,22	&	0,02	\\
827	&	0,21	&	0,02	&	0,22	&	0,02	\\
828	&	0,21	&	0,02	&	0,22	&	0,02	\\
\bottomrule
\end{tabular}
\caption{Changes of the FWHM (\%) evaluated before and after irradiation for BRDF and BTDF.}
\label{tab:FWHMtable}
\end{table}

\section{Discussion and Conclusions}
\label{sect:res}
We have introduced a method and reported experimental results aimed at characterizing the changes in the optical properties of glass due to radiation damage under conditions similar to those expected during space missions. 
Experimental results can be synthesized as follows: 
\begin{itemize}
\item The optical properties considered here are transmission and light scattering. Combining optical properties and chemical composition, it is possible to simulate collisions and damage for the glass samples. L.E.T. is a decreasing function of energy and it is not correlated with density. On the other hand, L.E.T. seems to be anti-correlated to the Lead fraction. Analyzing the energy absorbed for each chemical element, Lead appears to play a fundamental role, Calcium and Barium less but not negligible.  
\item Experimentation at KVI laboratories allowed to measure transmission before and after irradiation. By comparing transmission changes of space-compliant and no space-compliant glasses, it is possible to assess wether or not a glass can be used in space missions. We obtain the following results: 802, 805 and 807 are reliable in the whole range of wavelength; 804, 812, 813 and 814 only for space mission with an expected total dose of 10 krad. 
\item Some chemical elements influence transmission. With Lantanium issues appear in the wavelength range below 700 nm, as demonstrated by samples 808, 810 and 811. Fluorine seems to prduce oscillating spectran trends, as demonstrated by samples 819 and 828. 
\item Light scattered increases after irradiation for all samples. This suggests that radiation might affect the micro-roughness of the glass or even the internal structure. We have no insight into this issue. In any case, the angular spread of the scattered light increases with the absorbed dose. 
\end{itemize}

To achieve a better insght into several open points further measurements and simulations shall be done. Following the method proposed here, measurements could be extended to other glasses and transmission measurements could be done immediately after irradiation, in order to evaluate the effects of elastic properties, if any. Scattering measurements could be done also for glasses irradiated with 10 krad of total dose. This works then opens the way to different kinds of analyses, related to several issues encounetered in this work: 
\begin{itemize}
\item chemical composition analysis through X-diffraction before and after irradiation;
\item thermal effects analysis due to radiation and scattering;
\item study of the ultimate origin of scattering, that could be done through speckles analysis. 
\end{itemize}

\section*{Appendix}
\label{appendix}
Dose levels were computed considering specific orbits and mission length.
The properties of these orbits are simulated using SPENVIS that is ESA's Space ENVironment Information System: a web interface to model the space environment and its effects including cosmic rays, natural radiation belts, solar energetic particles, plasmas, gases, and "micro-particles".
Three different orbits were considered:
Geosynchronous orbits (GEO) are circular orbits around the Earth having a period of 24 hours. A geosynchronous orbit with an inclination of zero degrees is called a geostationary orbit. A spacecraft in an inclined geosynchronous orbit will appear to follow a regular figure-8 pattern in the sky. The  covering area of a geostationary satellite ideally extends up to an angle of $81\deg$ from the ground point directly under the satellite, that corresponds to something more than $40\%$ of the Earth's surface. 
In order to estimate the total dose for this kind of mission, protons and electrons' fluxes and fluences are evaluated following this model:
\begin{itemize}
\item AP-8 MAX and AE-8 MAX for Trapped Protons and Electrons respectively;
\item CRÉME-86 for the short-term solar particle flux considering from H to U;
\item KING solar proton model for the long-term solar particle fluence;
\item CRÉME-86 for GCR with 90\% worst case cosmic ray level.
\end{itemize}
Thanks to this type of simulation, it is possible to estimate the total dose for this type of orbit. Including a safety factor, a geostationary mission shows a total dose of about 10 krad.
Translunar orbit is a particular orbit around Earth and Moon, highly eccentric with an apogee around 360.000 km and a perigee close to 7.000 km from the Earth. This is due to particles (electrons in particular) trapped in the Van Hallen Belts for geostationary orbit. Thanks to this simulation, it is possible to estimate the total dose for this orbit. Including a safety factor, this mission shows a total dose of 10krad.
The last orbit we examined is a solar orbit. In particular, the orbit we considered has a minimum perihelion within 0.3 AU \cite{sol}. Another driver to go close to the Sun is the measurement of energetic particles, which should be made within one or two scattering mean free paths (typically 0.2 AU; Palmer 1982 \cite{palmer}) of their source in order to minimize propagation effects. 
Using, for example, the simulations of other scientific teams \cite{enge} , an analysis of L.E.T. and total dose can be done. The total dose analysis requires an assessment of the shielding provided by the spacecraft and equipment chassis. The L.E.T. spectrum is evaluated with CRÉME-96 model. The dose depth curve used in this analysis is taken from data \cite{ecss} and is the worst case at 0.28 AU. 
Using this type of simulation, it is possible to estimate the total dose for this orbit. Including a safety factor, this mission shows a total dose of 150 krad. This is the dose for the whole mission. Nevertheless, since the research was connected to a particular system integrated into the satellite, the glasses used in that instrument were analyzed for a limited total dose of 30 krad.
\pagebreak 

\bibliographystyle{spiejour} 
\bibliography{report}

\doublespacing
\listoffigures


\vspace{2ex}\noindent\textbf{Federica Simonetto} is PhD student at Università degli Studi di Milano. She received her BS and MS degrees in physics from the Università degli Studi di Milano in 2014 and 2017, respectively. Her current research interests include optical system, space procedures, hypergravity effect on proteins.

\vspace{1ex}\noindent\textbf{Matteo Marmonti} is Senior Optical Engineer at Optec S.p.A. He received his BS and MS degrees in physics from the Università degli Studi di Milano in 2007 and 2010, respectively. As optical engineer and expert in stray light analysis, he is constantly in contact with space agencies and centers operating in space field. He is also responsible of optical testing of telescopes and satellites and optical laboratory manager at Optec S.p.A.

\vspace{1ex}\noindent\textbf{Marco AC Potenza} leads the Instrumental Optics Laboratory at the Physics Department of the University of Milan and CIMAINA. His activity is focused at conceiving and developing novel optical techniques and instrumentation based on light scattering.
Since 2003 he works for developing space instrumentation for ESA.  He graduated in astroparticle physics in 1996. He got a Ph.D. In 1999 in non-linear optics, optical instrumentation and holography with the late Prof. Yury Denisyuk from St. Petersburg.\\

\begin{table}[ht]
\centering
\begin{tabular}{c|cccccc}
\toprule
Glass	&	$n_h$	&	$n_ g$	&	$n_{F'}	$	&	$n_F	$&	$n_e$	& $n_D$	\\
Code	&	405	&	436	&	480	&	486	&	546	&	589	\\
\midrule													
802	&	1,00849	&	1,00465	&	1,00393	&	1,00363	&	1,00205	&	1,00261	\\
804	&	0,89999	&	0,91894	&	0,93803	&	0,94033	&	0,95946	&	0,97175	\\
805	&	0,92051	&	0,93705	&	0,95747	&	0,96029	&	0,98297	&	0,99352	\\
806	&	1	&	0,95168	&	1,00065	&	1,00385	&	1,01435	&	1,01607	\\
807	&	0,94286	&	0,95886	&	0,97669	&	0,97902	&	0,99398	&	1,0014	\\
808	&	0,70147	&	0,7269	&	0,77138	&	0,77768	&	0,83345	&	0,86563	\\
810	&	0,69102	&	0,71935	&	0,76867	&	0,77559	&	0,83704	&	0,87155	\\
811	&	0,71456	&	0,73124	&	0,76913	&	0,77497	&	0,83297	&	0,86922	\\
812	&	0,8492	&	0,87148	&	0,89855	&	0,9015	&	0,92109	&	0,92869	\\
813	&	0,50002	&	0,54263	&	0,62899	&	0,6423	&	0,76578	&	0,84212	\\
814	&	0,82814	&	0,87067	&	0,91467	&	0,92014	&	0,95876	&	0,97572	\\
815	&	1,0033	&	1,02346	&	1,02573	&	1,0264	&	1,02675	&	1,02746	\\
816	&	0,78742	&	0,81114	&	0,82838	&	0,83017	&	0,83239	&	0,83365	\\
817	&	0,92895	&	0,9964	&	1,01778	&	1,0194	&	1,02427	&	1,02644	\\
818	&	0,78782	&	0,82335	&	0,85222	&	0,85543	&	0,87263	&	0,88464	\\
819	&	0,78134	&	0,75569	&	0,73241	&	0,73303	&	0,78738	&	0,86602	\\
820	&	0,78745	&	0,82131	&	0,85397	&	0,85843	&	0,907	&	0,94821	\\
821	&	0,99681	&	0,99993	&	0,99997	&	0,99987	&	0,99813	&	0,99657	\\
822	&	0,78996	&	0,84311	&	0,8871	&	0,89047	&	0,90459	&	0,90921	\\
823	&	0,74898	&	0,76881	&	0,79671	&	0,80005	&	0,83509	&	0,86281	\\
825	&	0,92895	&	0,87067	&	1,02573	&	0,77559	&	1,02427	&	0,84212	\\
826	&	1,00014	&	1,00825	&	1,00902	&	1,00911	&	1,00915	&	1,00999	\\
827	&	0,80101	&	0,86669	&	0,91862	&	0,92312	&	0,95356	&	0,96343	\\
828	&	0,64645	&	0,61811	&	0,60485	&	0,60588	&	0,67596	&	0,78476	\\
\bottomrule
\end{tabular}
\caption{Transmission variation evaluated before and after irradiation. Values about setup a. The second line contains wavelength values expressed in nanometer. Part 1.}
\label{tab:Ttable_a1}
\end{table}

\begin{table}[ht]
\centering
\begin{tabular}{c|cccccc}
\toprule
Glass	&	$n_{	632}$	&	$n_C	$ &	$n_r	$&	$n_s$ &	$n_t$	&	$n_{1060}$ \\
Code &	632	&	656	&	706	&	852	&	1014	&	1060 \\	
\midrule												
802 &	1,00142	&	1,00136	&	1,00117	&	1,00487	&	1,00515	&	0,98299	\\
804 &	0,981	&	0,98514	&	0,99101	&	1,00206	&	1,00003	&	0,9737	\\
805 &	1,00006	&	1,00374	&	1,01031	&	1,01871	&	1,01503	&	1,02245	\\
806 &	1,01503	&	1,01558	&	1,01649	&	1,02019	&	1,01491	&	0,99686	\\
807 &	1,00474	&	1,00724	&	1,01107	&	1,01551	&	1,00939	&	0,99168	\\
808 &	0,89491	&	0,91025	&	0,93823	&	0,98279	&	0,99542	&	1,00305	\\
810 &	0,9009	&	0,91573	&	0,94147	&	0,97951	&	0,99073	&	0,98847	\\
811 &	0,89832	&	0,91307	&	0,93845	&	0,9797	&	0,99467	&	0,98625	\\
812 &	0,93579	&	0,94087	&	0,95145	&	0,97503	&	0,98762	&	0,98357	\\
813 &	0,88553	&	0,89748	&	0,91124	&	0,97788	&	1,01532	&	1,05461	\\
814 &	0,98647	&	0,99393	&	1,00791	&	1,02897	&	1,02781	&	1,06212	\\
815 &	1,02689	&	1,02939	&	1,03535	&	1,04072	&	1,03103	&	1,06497	\\
816 &	0,84177	&	0,85163	&	0,87704	&	0,94079	&	0,98493	&	1,02724	\\
817 &	1,02658	&	1,02911	&	1,03542	&	1,04202	&	1,03171	&	1,06453	\\
818 &	0,90219	&	0,91731	&	0,95285	&	1,02147	&	1,02929	&	1,06753	\\
819 &	0,93592	&	0,96539	&	1,00323	&	1,03098	&	1,02567	&	1,06307	\\
820 &	0,98096	&	0,99479	&	1,01248	&	1,02664	&	1,02639	&	1,05786	\\
821 &	0,99553	&	0,99582	&	0,99722	&	0,9994	&	0,99925	&	1,01028	\\
822 &	0,91884	&	0,92753	&	0,94845	&	0,99032	&	0,99946	&	1,00115	\\
823 &	0,88774	&	0,90128	&	0,92614	&	0,9691	&	0,98802	&	0,9896	\\
825 &	0,88553	&	0,89748	&	0,91124	&	0,97788	&	1,01532	&	1,05461	\\
826 &	1,00946	&	1,01045	&	1,01388	&	1,01952	&	1,01878	&	1,05454	\\
827 &	0,97006	&	0,9741	&	0,98392	&	0,99409	&	0,99578	&	1,00882	\\
828 &	0,88354	&	0,92025	&	0,96113	&	0,98967	&	0,9957	&	0,99876	\\
\bottomrule
\end{tabular}
\caption{Transmission variation evaluated before and after irradiation. Values about setup a. The second line contains wavelength values expressed in nanometer. Part 2.}
\label{tab:Ttable_a2}
\end{table}

\begin{table}[ht]
\centering
\begin{tabular}{c|cccccc}
\toprule
Glass	&	$n_h$	&	$n_ g$	&	$n_{F'}	$	&	$n_F	$&	$n_e$	&$n_D$	\\
Code	&	405	&	436	&	480	&	486	&	546	&	589	\\
\midrule													
802	&	0,99321	&	0,98298	&	0,9821	&	0,98162	&	0,98028	&	0,98007	\\
804	&	0,88512	&	0,90709	&	0,93007	&	0,93259	&	0,95308	&	0,96557	\\
805	&	0,90997	&	0,92341	&	0,94455	&	0,94744	&	0,97063	&	0,98163	\\
806	&	1	&	0,95141	&	0,99371	&	0,99751	&	1,01032	&	1,01238	\\
807	&	0,93562	&	0,95163	&	0,97058	&	0,97299	&	0,98858	&	0,99649	\\
808	&	0,65169	&	0,67836	&	0,7282	&	0,7353	&	0,79872	&	0,83563	\\
810	&	0,63189	&	0,66421	&	0,72147	&	0,72948	&	0,80179	&	0,84236	\\
811	&	0,66325	&	0,68403	&	0,72014	&	0,7258	&	0,77883	&	0,80897	\\
812	&	0,80316	&	0,8426	&	0,88442	&	0,88969	&	0,93024	&	0,94758	\\
813	&	0,85056	&	0,87261	&	0,90267	&	0,90662	&	0,92775	&	0,93292	\\
814	&	0,79978	&	0,84715	&	0,89754	&	0,9039	&	0,94797	&	0,96604	\\
815	&	1,00044	&	1,02231	&	1,02539	&	1,02622	&	1,02691	&	1,02757	\\
816	&	0,75369	&	0,77975	&	0,79922	&	0,80129	&	0,8033	&	0,80402	\\
817	&	0,90199	&	0,98328	&	1,0113	&	1,01337	&	1,02066	&	1,02335	\\
818	&	0,74144	&	0,78093	&	0,81486	&	0,81861	&	0,8387	&	0,85177	\\
819	&	0,72887	&	0,70101	&	0,67703	&	0,67794	&	0,74061	&	0,83161	\\
820	&	0,73203	&	0,76762	&	0,80561	&	0,81056	&	0,86667	&	0,91348	\\
821	&	0,99142	&	0,9952	&	0,99652	&	0,9965	&	0,99563	&	0,99473	\\
822	&	0,7407	&	0,80318	&	0,8563	&	0,86047	&	0,87899	&	0,88577	\\
823	&	0,70138	&	0,7228	&	0,75396	&	0,75781	&	0,79797	&	0,83	\\
825	&	1,00044	&	1,02231	&	1,02539	&	1,02622	&	1,02691	&	1,02757	\\
826	&	0,99575	&	1,00132	&	1,00151	&	1,00145	&	1,00013	&	0,99929	\\
827	&	0,72775	&	0,79726	&	0,86356	&	0,87011	&	0,91219	&	0,92736	\\
828	&	0,58081	&	0,5494	&	0,53544	&	0,53668	&	0,61537	&	0,74058	\\
\bottomrule
\end{tabular}
\caption{Transmission variation evaluated before and after irradiation. Values about setup b. The second line contains wavelength values expressed in nanometer. Part 1.}
\label{tab:Ttable_b1}
\end{table}

\begin{table}[ht]
\centering
\begin{tabular}{c|cccccc}
\toprule
Glass	&	$n_{	632}$	&	$n_C	$ &	$n_r	$&	$n_s$ &	$n_t$	&	$n_{1060}$ \\
Code	&	632	&	656	&	706	&	852	&	1014	&	1060 \\	
\midrule													
802	&	0,98018	&	0,9793	&	0,97696	&	0,97612	&	0,97996	&	0,94497	\\
804	&	0,97508	&	0,97877	&	0,98355	&	0,99386	&	0,99357	&	0,96275	\\
805	&	0,9891	&	0,99191	&	0,99624	&	1,00394	&	1,00353	&	0,99046	\\
806	&	1,01149	&	1,01158	&	1,0112	&	1,0136	&	1,00955	&	0,98539	\\
807	&	0,99999	&	1,002	&	1,00425	&	1,00564	&	0,99804	&	0,97431	\\
808	&	0,86931	&	0,88635	&	0,91598	&	0,95688	&	0,96682	&	0,96369	\\
810	&	0,87522	&	0,89081	&	0,91567	&	0,94606	&	0,95252	&	0,9443	\\
811	&	0,83124	&	0,84145	&	0,85592	&	0,8713	&	0,87527	&	0,86526	\\
812	&	0,95867	&	0,96451	&	0,97413	&	0,99236	&	1,00089	&	0,99154	\\
813	&	0,93379	&	0,93579	&	0,93791	&	0,93038	&	0,92145	&	0,95171	\\
814	&	0,97612	&	0,98239	&	0,9923	&	1,0005	&	0,99337	&	1,0246	\\
815	&	1,02697	&	1,02931	&	1,03479	&	1,04066	&	1,03163	&	1,06334	\\
816	&	0,81294	&	0,82361	&	0,85055	&	0,91746	&	0,96663	&	1,00874	\\
817	&	1,02401	&	1,02691	&	1,03314	&	1,0392	&	1,02974	&	1,06416	\\
818	&	0,8725	&	0,89003	&	0,93104	&	1,00893	&	1,01808	&	1,05438	\\
819	&	0,91385	&	0,94863	&	0,9925	&	1,02814	&	1,02478	&	1,06166	\\
820	&	0,95206	&	0,9675	&	0,98656	&	1,00073	&	1,00173	&	1,02267	\\
821	&	0,99384	&	0,99431	&	0,99535	&	0,99825	&	0,9989	&	1,00527	\\
822	&	0,89811	&	0,90869	&	0,93379	&	0,98574	&	0,99765	&	1,00012	\\
823	&	0,85944	&	0,87547	&	0,90469	&	0,95663	&	0,9815	&	0,98566	\\
825	&	1,02697	&	1,02931	&	1,03479	&	1,04066	&	1,03163	&	1,06334	\\
826	&	0,99847	&	0,99906	&	1,00142	&	1,00588	&	1,00629	&	1,02708	\\
827	&	0,93988	&	0,94837	&	0,96754	&	0,99051	&	0,99454	&	1,00348	\\
828	&	0,85809	&	0,90263	&	0,95217	&	0,98619	&	0,99385	&	0,99907	\\
\bottomrule
\end{tabular}
\caption{Transmission variation evaluated before and after irradiation. Values about setup b. The second line contains wavelength values expressed in nanometer. Part 2.}
\label{tab:Ttable_b2}
\end{table}

\begin{table}[ht]
\centering
\begin{tabular}{c|cccccccccccc}
\toprule
 Glass	&	$n_h$	&	$n_ g$	&	$n_{F'}	$	&	$n_F	$&	$n_e$	&$n_D$	\\
Code	&	405	&	436	&	480	&	486	&	546	&	589	\\
\midrule													
802	&	1,00155	&	1,002	&	1,00034	&	1,00031	&	0,99743	&	0,99596	\\
804	&	0,95289	&	0,96358	&	0,97311	&	0,97463	&	0,98411	&	0,98998	\\
805	&	0,94383	&	0,95596	&	0,96966	&	0,97179	&	0,9855	&	0,99161	\\
806	&	1	&	0,95995	&	1,00645	&	1,00803	&	1,01003	&	1,00951	\\
807	&	0,95361	&	0,97098	&	0,98407	&	0,98582	&	0,9962	&	1,00038	\\
808	&	0,87303	&	0,8861	&	0,90596	&	0,90872	&	0,93096	&	0,94318	\\
810	&	0,87143	&	0,88601	&	0,9077	&	0,91068	&	0,93492	&	0,94786	\\
811	&	0,87256	&	0,88246	&	0,90051	&	0,90329	&	0,92836	&	0,9436	\\
812	&	0,93616	&	0,95479	&	0,96907	&	0,97121	&	0,98304	&	0,98779	\\
813	&	0,94094	&	0,96016	&	0,97539	&	0,97811	&	0,99321	&	0,99957	\\
814	&	0,92704	&	0,95147	&	0,97137	&	0,97409	&	0,99154	&	0,99914	\\
815	&	0,99689	&	1,01391	&	1,01655	&	1,01745	&	1,01894	&	1,02017	\\
816	&	0,9132	&	0,92919	&	0,93779	&	0,93909	&	0,94098	&	0,94229	\\
817	&	0,96986	&	1,00369	&	1,01316	&	1,01431	&	1,01761	&	1,01891	\\
818	&	0,92186	&	0,9415	&	0,95292	&	0,95461	&	0,96126	&	0,96564	\\
819	&	0,91435	&	0,90773	&	0,8973	&	0,89813	&	0,92272	&	0,95586	\\
820	&	0,90126	&	0,91612	&	0,93049	&	0,93228	&	0,95062	&	0,96456	\\
821	&	0,99344	&	0,99548	&	0,99559	&	0,9956	&	0,99318	&	0,99159	\\
822	&	0,91558	&	0,9365	&	0,95282	&	0,95408	&	0,95719	&	0,95706	\\
823	&	0,90196	&	0,91154	&	0,92373	&	0,92527	&	0,93813	&	0,94754	\\
825	&	0,9132	&	0,92919	&	0,93779	&	0,93909	&	0,94098	&	0,94229	\\
826	&	0,98925	&	0,99421	&	0,99476	&	0,99476	&	0,99362	&	0,99284	\\
827	&	0,90649	&	0,93284	&	0,95496	&	0,95705	&	0,96918	&	0,97328	\\
828	&	0,85053	&	0,83862	&	0,83246	&	0,83288	&	0,86498	&	0,91174	\\
\bottomrule
\end{tabular}
\caption{Transmission variation evaluated before and after irradiation. Values about setup c. The second line contains wavelength values expressed in nanometer. Part 1.}
\label{tab:Ttable_c1}
\end{table}

\begin{table}[ht]
\centering
\begin{tabular}{c|cccccccccccc}
\toprule
Glass		&	$n_{	632}$	&	$n_C	$ &	$n_r	$&	$n_s$ &	$n_t$	&	$n_{1060}$ \\
Code	&	632	&	656	&	706	&	852	&	1014	&	1060 \\	
\midrule													
802	&	0,99452	&	0,99489	&	0,9958	&	0,9954	&	0,99516	&	0,99643	\\
804	&	0,99409	&	0,99709	&	1,00212	&	1,00936	&	1,00548	&	1,00754	\\
805	&	0,99563	&	0,99802	&	1,00272	&	1,00936	&	1,00657	&	1,01878	\\
806	&	1,00817	&	1,00886	&	1,01027	&	1,01169	&	1,00547	&	1,00646	\\
807	&	1,00285	&	1,00529	&	1,0099	&	1,0188	&	1,01883	&	1,05452	\\
808	&	0,95462	&	0,96082	&	0,97223	&	0,99236	&	0,99853	&	1,01218	\\
810	&	0,95896	&	0,96477	&	0,97492	&	0,99118	&	0,99578	&	1,00412	\\
811	&	0,95432	&	0,95963	&	0,9677	&	0,97577	&	0,97872	&	0,98635	\\
812	&	0,99141	&	0,99608	&	1,00627	&	1,01971	&	1,01728	&	1,06721	\\
813	&	1,00353	&	1,00799	&	1,01693	&	1,02708	&	1,02145	&	1,06731	\\
814	&	1,00383	&	1,00841	&	1,01782	&	1,02967	&	1,02562	&	1,07482	\\
815	&	1,02038	&	1,02298	&	1,02892	&	1,03608	&	1,02894	&	1,07512	\\
816	&	0,94592	&	0,95163	&	0,96571	&	0,99609	&	1,00963	&	1,05907	\\
817	&	1,01942	&	1,02215	&	1,02837	&	1,03294	&	1,02459	&	1,07217	\\
818	&	0,97225	&	0,97958	&	0,99671	&	1,0239	&	1,0218	&	1,07055	\\
819	&	0,98407	&	0,99713	&	1,0153	&	1,0279	&	1,02158	&	1,07008	\\
820	&	0,97658	&	0,98159	&	0,98867	&	0,99472	&	0,99665	&	1,02081	\\
821	&	0,99053	&	0,99085	&	0,99159	&	0,99305	&	0,9944	&	1,00557	\\
822	&	0,95967	&	0,96307	&	0,97117	&	0,98478	&	0,98829	&	0,99736	\\
823	&	0,95623	&	0,96143	&	0,97085	&	0,98458	&	0,991	&	1,00028	\\
825	&	0,94592	&	0,95163	&	0,96571	&	0,99609	&	1,00963	&	1,05907	\\
826	&	0,9923	&	0,99248	&	0,9944	&	0,99857	&	1,00116	&	1,02394	\\
827	&	0,97647	&	0,9791	&	0,98564	&	0,99441	&	0,99764	&	1,01053	\\
828	&	0,95096	&	0,9653	&	0,98135	&	0,99278	&	0,99801	&	1,01011	\\
\bottomrule
\end{tabular}
\caption{Transmission variation evaluated before and after irradiation. Values about setup c. The second line contains wavelength values expressed in nanometer. Part 2.}
\label{tab:Ttable_c2}
\end{table}

\begin{figure}[h]
\centering
\caption{\textbf{Glasses 802 and 806}: scattering before and after irradiation. (a): 802 BRDF functions, before and after irradiation. (b): 802 BTDF functions, before and after irradiation. (c): 806 BRDF functions, before and after irradiation. (d): 806 BTDF functions, before and after irradiation.}
\includegraphics[scale=0.6]{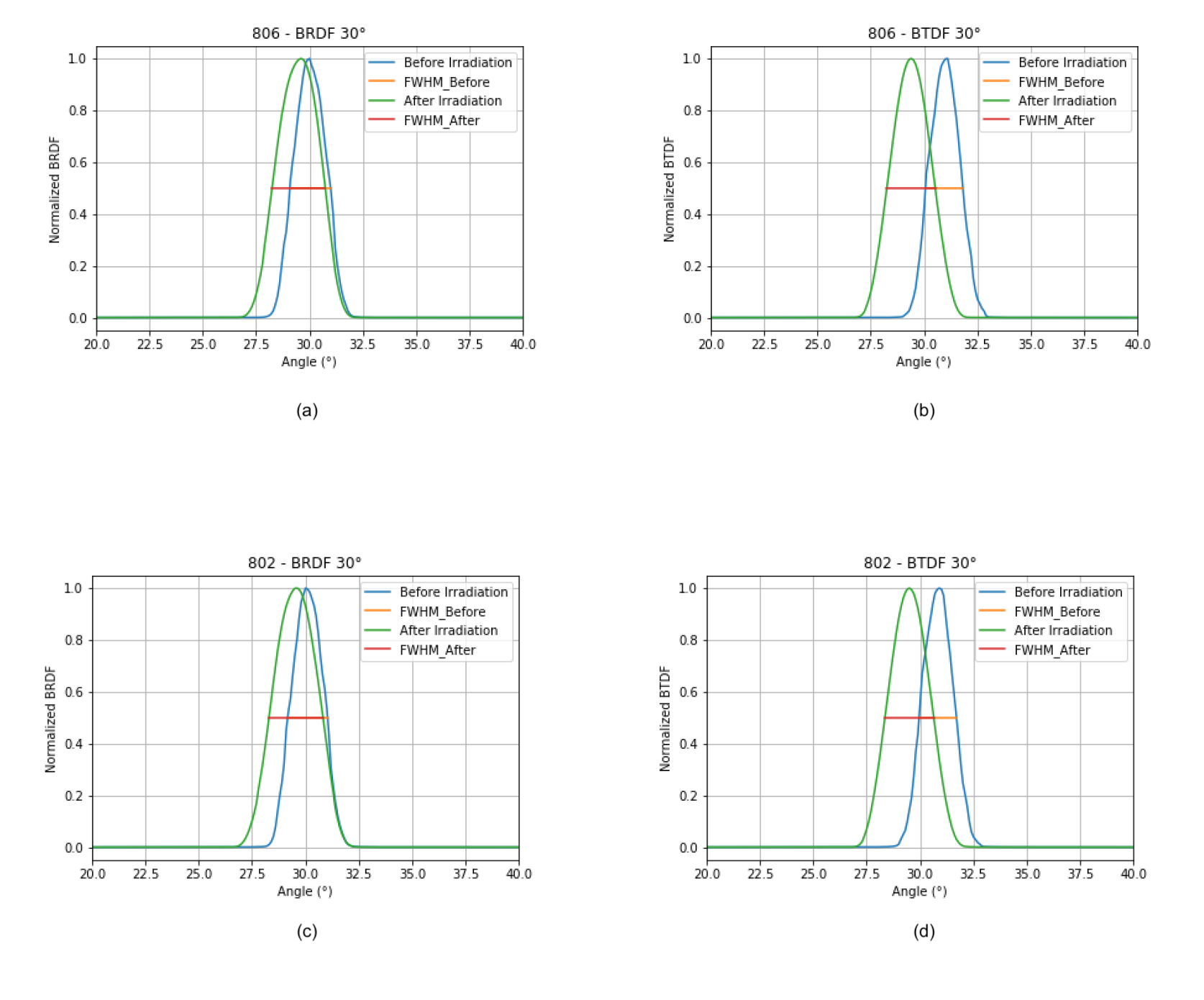}
\label{fig:806}
\end{figure}

\begin{figure}[h]
\centering
\caption{\textbf{Glasses 810 and 820}: scattering before and after irradiation. (a): 810 BRDF functions, before and after irradiation. (b): 810 BTDF functions, before and after irradiation. (c): 820 BRDF functions, before and after irradiation. (d): 820 BTDF functions, before and after irradiation.}
\includegraphics[scale=0.6]{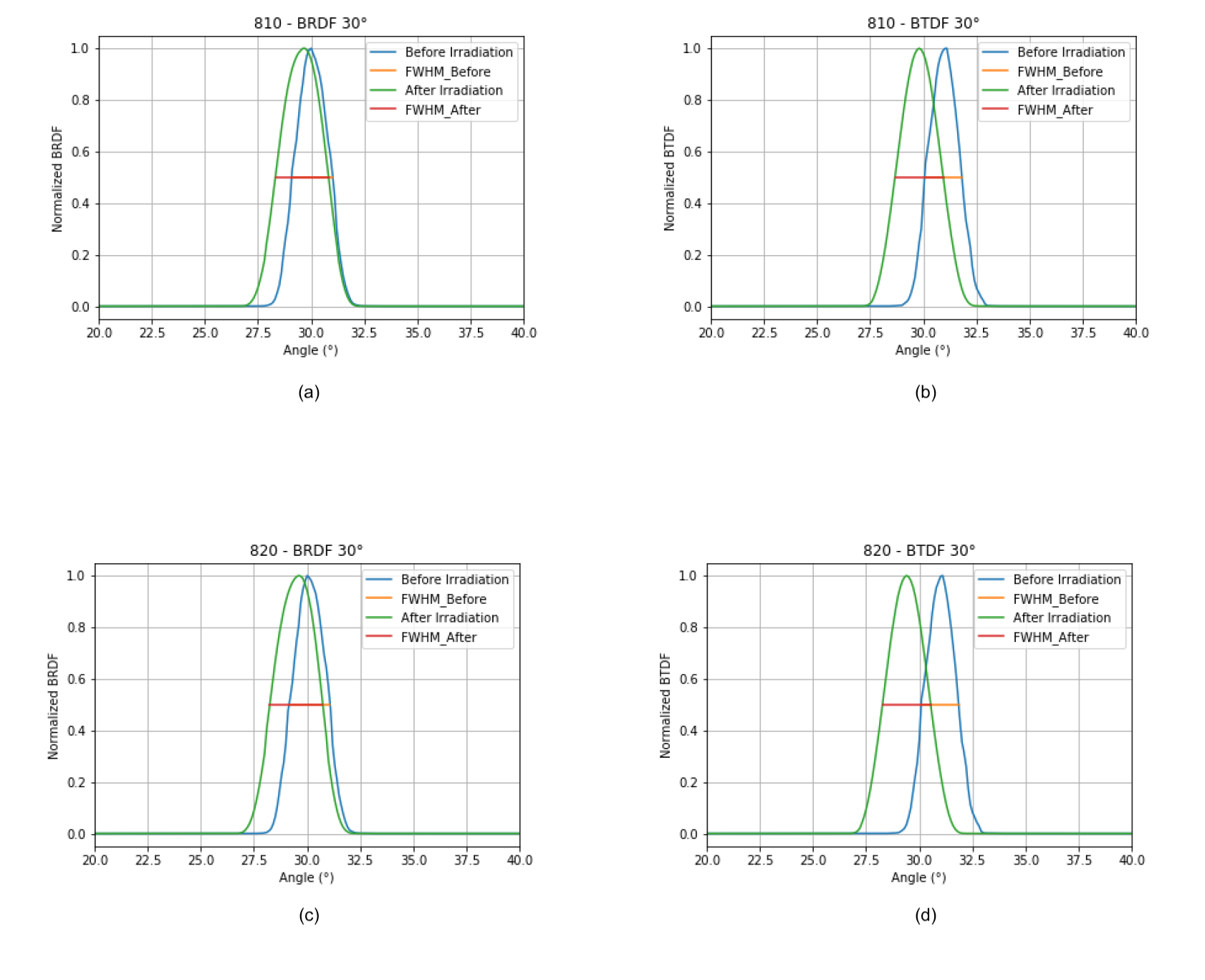}
\label{fig:810}
\end{figure}

\end{spacing}
\end{document}